# Communication of Statistics and Evidence in Times of Crisis

Claudia R. Schneider[1,2], John R. Kerr[3], Sarah Dryhurst[4,2], John A. D. Aston[5]

[1]Department of Psychology, University of Cambridge, Cambridge, United Kingdom

[2]Winton Centre for Risk and Evidence Communication, Department of Pure Mathematics and Mathematical Statistics, University of Cambridge, Cambridge, United Kingdom

[3]Department of Public Health, University of Otago, Wellington, New Zealand

[4]Institute for Risk and Disaster Reduction, University College London, London, United Kingdom

[5]Statistical Laboratory, Department of Pure Mathematics and Mathematical Statistics, University of Cambridge, Cambridge, United Kingdom

CS: cs2025@cam.ac.uk
JK: john.kerr@otago.ac.nz
SD: s.dryhurst@ucl.ac.uk
JA: j.aston@statslab.cam.ac.uk

CS: http://orcid.org/0000-0002-6612-5186
JK: http://orcid.org/0000-0002-6606-5507
SD: http://orcid.org/0000-0002-7772-8492
JA: https://orcid.org/0000-0002-3770-9342

Corresponding authors contact information:
John A. D. Aston, Statistical Laboratory, Department of Pure Mathematics and Mathematical Statistics, University of Cambridge, Cambridge, United Kingdom,
j.aston@statslab.cam.ac.uk
Claudia R. Schneider, Department of Psychology, University of Cambridge, Cambridge, United Kingdom
cs2025@cam.ac.uk

## Abstract

This review provides an overview of concepts relating to the communication of statistical and empirical evidence in times of crisis, with a special focus on COVID-19. In it, we consider topics relating both to the communication of numbers – such as the role of format, context, comparisons, and visualization – and the communication of evidence more broadly – such as evidence quality, the influence of changes in available evidence, transparency, and repeated decision making. A central focus is on the communication of the inherent uncertainties in statistical analysis, especially in rapidly changing informational environments during crises. We present relevant literature on these topics and draw connections to the communication of statistics and empirical evidence during the COVID-19 pandemic and beyond. We finish by suggesting some considerations for those faced with communicating statistics and evidence in times of crisis.

# 1. Introduction

When the COVID-19 pandemic suddenly struck, scientific evidence, statistics, and graphs became prime-time television viewing. Policy-makers and the public suddenly found themselves discussing epidemiological and scientific concepts, often expressed as numbers, in a way that would have been unforeseeable just a few weeks before. However, as Spiegelhalter said in his 2017 review "numbers do not speak for themselves" (Spiegelhalter, 2017, p. 37): they are in the hands of those communicating them, and with the outbreak of the pandemic, many found themselves suddenly holding the responsibility of risk and evidence communication without experience.

Communicating empirical evidence is not just about trying to get across a 'number' - the aim instead is to communicate the understanding of that number: what it actually represents. For example, what is important about communicating an R number of 4 is not that the audience come away remembering the number '4'. It's that they understand that it means that we believe the epidemic is in a period of fast, exponential growth; that they understand the real consequences of that; and also that they understand the uncertainties around that number: how sure are we that the R number is '4'; how sure are we that it is above 1? This means that evidence communicators need to understand how people respond to and interpret different presentations of numbers and uncertainties.

In this review we attempt to shed light on some of the most important concepts in this field; focusing on what we feel is some of the most relevant and significant literature. We see this review as a starting point for further reading, as this is a vast area, and in many places we simply scratch the surface of what can be useful in practice. While we hope that we have surveyed the literature in general, we've deliberately chosen illustrative examples from COVID-19 communication that we were involved in at the Winton Centre for Risk and Evidence Communication as opposed to subjectively calling out examples from the pandemic as most communication during the crisis was not subjected to empirical evaluation.

We also note that with any type of communication, keeping the audience in mind is crucially important as communications should to be adapted to their needs and knowledge levels. For instance, communication of statistics and empirical evidence to the public might look different from communication to policy-makers and decision makers, and yet different again to communication between experts. There are many common principles, but details might differ. While we hope to provide some useful insights that could support communicators in times of crisis we also highlight the fundamental importance of testing and evaluating communication in its specific context and with its specific audience where possible.

## 1.1. Statistics and empirical evidence in times of crisis

The mantra is often that information is power, but the converse is almost certainly more apposite in a time of crisis – a lack of information makes people feel powerless. When SARS-CoV-2 first appeared in Wuhan in late 2019, and then spread around the world, there was a lack of available evidence to allow decision makers and the general public to know and understand the risks and make decisions for their own or others' safety.

In any kind of crisis, the role of evidence and information is vital and in particular there is almost always a dynamic nature to the evidence as more is gathered or as the understanding changes over time. People need to make decisions, and indeed to know when no action becomes itself an active decision. This requires evidence, and in a public health crisis such as the COVID-19 pandemic, much of the available evidence is quantifiable, leading to a range of statistics which need to be communicated. However, quantifiable doesn't mean certain, hence uncertainties about the produced statistics and the underlying evidence need to be communicated alongside. When evidence did start to emerge around COVID-19, it often came in the form of numbers and concepts unfamiliar to all but those with considerable training in epidemiology. While many non-experts now know about R numbers, doubling times and exponential growth, this was not the case in early 2020. These are not statistics that many people use on a daily basis, and require context to make sense of them.

However, a few key evidential facts do not lead to a complete understanding. As more evidence emerges, it needs to be communicated. This requires statisticians and analysts to say what is known, and what isn't, and indeed the levels of uncertainty and quality in the evidence being presented. Important features might be whether data is projected from a model, or whether data arises from a survey, or explaining the idiosyncrasies of administrative data (such as day-of-the-week effects). These factors are known to experts, but also need to be taken into account by those making decisions, be those decisions of national importance or personal, individual ones. Meaning needs to be taken from heterogeneous sources of differing quality, while not implying that there is so much uncertainty that we really know nothing at all. It is here that statistical and empirical evidence communication more broadly becomes vitally important.

At the outset of the COVID-19 pandemic, statistical (and more widely, quantitative evidence) communicators were suddenly in demand and one of the few sources of information in a world that seemed to be completely different from even a few days before. This placed considerable pressure on those performing this role, as they aimed to allow evidence to get to those who needed it, in a manner that was designed to inform but not persuade.

## 1.2. Why the communication of empirical evidence and its psychological perspectives matter

The way in which empirical evidence and statistics are communicated matters for how people perceive the information, to what extent they trust it, how they use it in their decision making and ultimately how it affects their behaviour (Gustafson & Rice, 2020; Howe et al., 2019; Johnson & Slovic, 1995; Recchia et al., 2021; van der Bles et al., 2020). Hence, we cannot consider statistics in numeric isolation, we must consider their psychological and behavioural effects on audiences when communicated. Understanding the various factors and mechanisms at play can help to design communication which avoids unintended effects.

A vital consideration is the aim of the communication, distinguishing between communication that aims to persuade the audience toward a certain course of action, and communication that aims to support decision making without guiding toward a particular outcome (Blastland et al., 2020).

Communication with the aim to persuade might involve strategies to for instance "engender maximum support and participation" (Hyland-Wood et al., 2021, p. 1). Communication with the aim to inform will not focus on a particular decision outcome, but rather on the process of ensuring that informed decision making is possible, by for instance providing a balanced account of harms and benefits, disclosing uncertainties and evidence quality, and pre-empting misunderstandings – a process termed 'evidence communication' (Blastland et al., 2020; Kerr et al., 2022). Audience understanding of the communicated information is critically important in evidence communication, where 'understanding' is about the concept behind a number, not just the number itself: to be able to weigh it up in decision-making. That makes the aims – and hence measures of success – of evidence communication less easily assessed. Typical measures of success for a persuasive, message-based communication might be relatively straight forward, tapping into things like 'how many people took/said they would take a particular action after reading it?'. For evidence communication, success is measured by asking questions which assess a person's understanding of the concept and gist of what is being communicated – alongside measures of how helpful it was in their decision-making, and how trustworthy they found it. Trust and trustworthiness are key because they are known to play an important role for how people use and act on information (Cologna & Siegrist, 2020; Schneider et al., 2022; Taufique et al., 2017).

'Should I be informing or persuading?' is something a communicator should consider every time they are designing a piece of communication. Things that they might want to consider are:
(1) Does the audience have time to read and consider evidence before making a decision (e.g. an Operational Earthquake Forecast, giving the likelihood of a damaging earthquake in a particular area within the next 7 days), or is this an acute emergency (e.g. Earthquake Early Warning, where they may be alerted a matter of seconds before an earthquake hits)?
(2) Has the audience consented – either explicitly or implicitly – to being persuaded and to the means of persuasion intended to be employed?
(3) Is there an ethical imperative to inform rather than persuade (e.g. in doctor-patient interactions and shared decision-making)? When might a persuasive approach be ethically defensible? (e.g. might evacuation mandates in case of natural disasters be such a case?)
Table 1 provides an overview of differing aims of communication to inform versus persuade.

**Table 1.** Aims of informing versus persuasive communication approaches.

|  | **Informing approach** | **Persuasive approach** |
|---|---|---|
| Overall aim is to... | inform | persuade |
| Make people... | understand | believe |
| Outcome measure: | better informed | changed behaviour |
| Communication tool: | information | a message |

# 2. Communicating numbers

## 2.1. The role of affect and social context

Humans are not rational machines: our judgments and decisions are shaped by our emotional (affective) reactions to, and processing of, information (Finucane et al., 2000; Slovic & Peters, 2006). This means that statistically equivalent probabilities, for example – even if they have a similar impact - might not feel equivalent to us, and not everyone will feel the same way about every risk. Risk is not just a number, it is a feeling as well (Slovic et al., 2005; Slovic & Peters, 2006).

This may partially explain why people had differing reactions to the threat of the SARS-CoV-2 virus. They may have perceived the risk differently – due to their differing experiences relating to the virus (e.g. had they known someone personally who died), or the differing impact that being ill might have on each of them (e.g. whether they had caring or professional responsibilities that would be severely affected if they became ill).

Research has shown that information that evokes emotions can grab and hold our attention and influence decision making (Ohman et al., 2001). While an emotional reaction may help us to attend to important information, it may come with the danger of emotional numbing, issue fatigue, and disengagement when used as a means of communication – such as with the issue of climate change (O'Neill & Nicholson-Cole, 2009; Pidgeon, 2012; Stoknes, 2014). Similarly, content that is more relevant to us may hold our attention more and make us more likely to seek further information (Falbén et al., 2020; Sui & Rotshtein, 2019).

COVID-19 is a highly emotive topic and of relevance to all of us personally. Therefore, it is not surprising that people actively sought more information. For example, one UK national survey estimated that more than a third of adults accessed the Public Health England data dashboard for COVID-19 information, and 15% visited the Office for National Statistics website (Sense about Science, 2022). Crises are a unique opportunity – and challenge – for statisticians to communicate to an audience who is interested and ready to listen.

Direct experience with the virus played an important role for risk perception during the pandemic both longitudinally in the UK (Schneider et al., 2021) as well as internationally in several countries around the world (Dryhurst et al., 2020). There was also a social context to COVID-19 risk perception; in particular, social amplification of risk (Kasperson et al., 1988; Pidgeon, 2012). Receiving information about COVID-19 from friends and family was more predictive of risk perceptions of the virus than measures of objective knowledge (such as how much scientists knew about the virus at the time). This is not surprising, as social amplification is known to play a role in high dread and unfamiliar contexts, such as the COVID-19 pandemic (Haas, 2020).

These findings highlight the importance of taking into account affective and social aspects when thinking about the communication of statistics and empirical evidence.

## 2.2. A number is not just a number, format matters

Numbers do not exist in a vacuum. How they are presented, both in terms of the format and the words around them, can have an impact on the audience and the choices they make based on that information.

*Percentages and expected frequencies*

Would you rather take a treatment with 5% chance of death or 1 in 20 chance of death? What about if 5 in every 100 people who took it died? Most statistical readers will have noted that those risks are statistically equivalent. However, the way they are presented creates different perceptions.

Generally, people perceive probabilities presented as percentages (e.g. '5% chance of X') as less likely or risky than expected frequencies (e.g. '1 in 20 people will experience X') (Freeman et al., 2021). And they perceive expected frequencies with a smaller numerator as less risky than those with a larger numerator (e.g. '1 in 20' vs. '10 in 200') (Denes-Raj et al., 1995). Expected frequencies may be more useful than percentages when referring to very low probabilities ($< 1\%$) where percentages would require decimals (Spiegelhalter, 2017).

Communicators should avoid comparing expected frequencies with differing denominators ('1 in x' format). Readers must perform mental gymnastics to compare the magnitude of, say, a 1 in 500 chance with a 1 in 700 chance, and may erroneously use the size of the denominator ($700 > 500$) as a guide for approximating relative probabilities--perversely perceiving less likely outcomes as more likely (Cuite et al., 2008; Woloshin et al., 2000).

One way to mitigate the unintended consequences of format choices is to present statistics in multiple formats e.g., 'this intervention has a 5% success rate, that means it will work for five out of every 100 people'.

*Framing*

Choices about whether statistics refer to death or survival, success or failure can affect audiences' perceptions and decisions. These, of course, were statistics that became important during COVID-19. For example, people are more likely to opt for a treatment when it is described as having a 95% survival rate rather than a 5% fatality rate (Moxey et al., 2003). When the information is framed positively (i.e. in terms of benefits such as survival), people may be more likely to choose the treatment because they want to increase their chances of survival. In contrast, when the information is framed negatively, people may be more likely to avoid the treatment because they want to minimize the risk of death. In a study looking at the communication of personalized risk from COVID-19 people who received negatively framed information ('We'd expect 2% of people with this result to die if they got COVID-19') perceived the information as more uncertain compared to those who received it with a positive frame ('We'd expect 98% of people with this result to survive if they got COVID-19') (Freeman et al., 2021). One clear way to avoid the effects of such framing is to provide the statistic from both sides (Gigerenzer, 2014; Spiegelhalter, 2017). Visual icon arrays (section 2.7) are effective at doing this as they display (for example) the proportion of expected deaths *and* survivors in a single graphic.

*Relative vs absolute*

Many statistics can be communicated in relative (e.g. x% increase) or absolute (e.g. increase from x% to x%) terms, such as changes over time or increases or reductions in risk. As clearly shown in the XKCD webcomic (Figure 1), relative statistics can be misleading without information on absolute numbers.

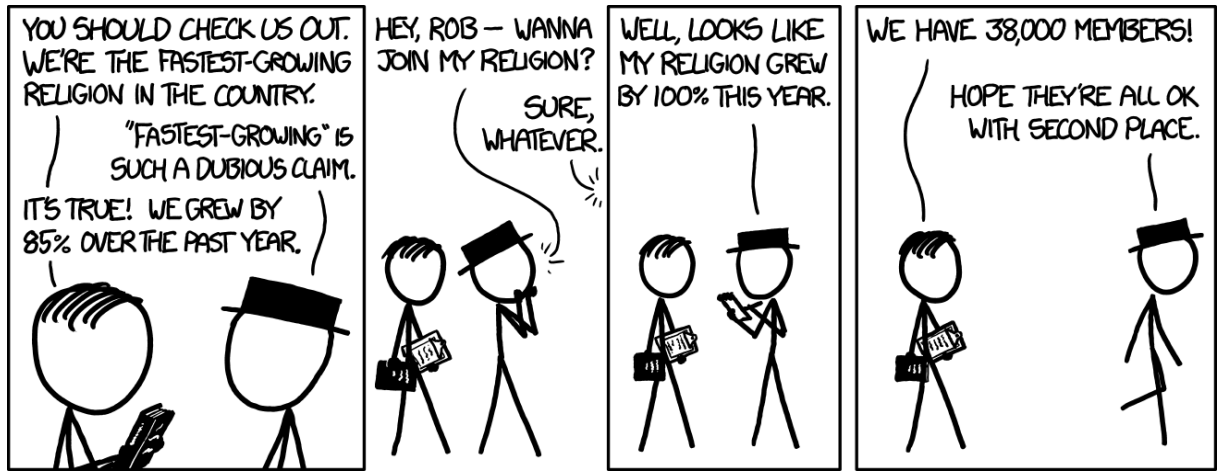

**Figure 1.** XKCD webcomic. The webcomic highlights the perils of relative statistics. Source: xkcd.com/1102; Republished under a CC BY-NC 2.5 Creative Commons licence.

While relative risks and statistics can convey important information, they should be reported alongside baselines or absolute changes.

## 2.3. Why communicate numbers and not words?

There is a temptation to express numerical values in non-numeric verbal terms like 'overwhelming majority' or 'very likely'. Unfortunately, these words can mean very different things to different people (Büchter et al., 2014). For example, in the absence of numeric information, people over estimate the risks of COVID-19 vaccine side effects when they are described using standardised regulatory terms such as 'common' or 'very rare' (Shoots-Reinhard et al., 2022).

Scales which align verbal terms with numeric probabilities can be found in the domains of food safety, environmental, and national security risk communication (reviewed in Teigen, 2022). One example is the Intergovernmental Panel on Climate Change's (IPCC) recent report which discusses the probability of future events in terms such as 'unlikely' and 'almost certain' (IPCC, 2022). In a footnote in the report, these terms are defined as corresponding to specific risks as a percentage range. For example, 'unlikely' means a 0-33% chance, while 'almost certain' means 99-100% chance.

Providing such definitions is an improvement over words alone but an international evaluation of the IPCC scale noted that the risks defined by verbal terms don't necessarily align with the public's perceptions of what those words mean (Budescu et al., 2012). Individuals' interpretations of IPCC risk statements were more accurate when the words and numbers were presented together *every time* in the texts, rather than simply in a separate reference table (Budescu et al., 2012).

While there is ongoing debate over which words do a better job of representing different numeric risks in different contexts (e.g. Mandel et al., 2021; Mandel & Irwin, 2021), our core advice remains: use words and numbers to communicate statistics, rather than words alone. The choice of using a numeric versus a verbal format when communicating statistics and empirical information does not only matter for the understanding of the underlying numeric information, but also for outcomes in the context of the communication of uncertainty (see section 2.6). While we presented relevant research and literature on the communication of numbers, it might be useful to illustrate how users of this review, including decision makers in the real world or statisticians, might think about the presented topics and approach a crystallization of them to give a practical perspective. While we are not offering an exhaustive list and not covering all topics discussed in this review, Table 2 gives examples of the issues we have presented above in a direct issue-mitigation-(hypothetical) example format. This shows that many issues can be framed relatively simply and that the mitigations can be relatively easily implemented.

**Table 2.** Issue, mitigation, and (hypothetical) example illustrations for three topics on the communication of numbers.

| Issue | Mitigation | Example |
|---|---|---|
| People process percentages and expected frequencies in different ways | Present numbers in multiple formats, making sure you use frequencies with the same denominator | 'Intervention A has a 5% success rate, that means it will work for five out of every 100 people; this is higher than the success rate of intervention B (3%, or 3 out of every 100 people)' |
| The way the same risk is framed has an impact | Avoid the effects of such framing by providing the statistic from both sides | 'We'd expect 98% of people to survive after the treatment, which means we'd expect 2% of people to die after the treatment.' |
| Quantitative-sounding but non-specific words mean different things to different people | Use words and numbers to communicate statistics, rather than words alone | 'The side effect of this treatment is very common (21% likelihood).'* |

* Example adapted from Shoots-Reinhard et al., 2022

## 2.4. Context and contextual language matter

A number on its own doesn't mean much to people. Numerical language helps to convey quantities with precision – but it doesn't help to interpret what that magnitude actually means. Without experience, the effect of an R number of 2, 1 or 0.5 is difficult to imagine.

Where people are weighing up options in a decision, the most useful way to give context might be the expected impact in each case so that they can judge how much and what sort of difference there is between the outcomes. For example, for a patient to decide whether to take a medical treatment or not, they need to know the likelihood of the benefit with and without the treatment, and the likelihood of harms with and without the treatment. These comparisons give context, as for instance shown in the AstraZeneca COVID-19 vaccine graphic which was produced by the Winton Centre for Risk and Evidence Communication (see Section 2.7.).

For hazards such as earthquakes or floods, where communicators might be trying to convey the severity of a forecast in a period of elevated risk, it might be useful to give a comparison to the "normal" level of risk for the forecast location, or comparison with the level of risk in other locations or times which might be familiar to the audience. For instance, before the infamous, destructive 2009 L'Aquila earthquake hit only the low absolute risk was communicated, failing to acknowledge however that the likelihood was about 100 to 1000 times higher than the "normal" risk for that region which would have provided important context to people (Fountain, 2011). Communications in such cases could be in the form of a relative risk (not on their own, as mentioned in 2.2, but alongside the absolute risk), or a series of absolute risks, perhaps portrayed graphically on a risk ladder (Savadori et al., 2022).

Linked to the role of context, it is also important to highlight effects such as anchoring (Tversky & Kahneman, 1974). Anchoring refers to the concept that our perceptions of magnitude can be affected by numbers that we've just seen. So once a number or a statistic is communicated, people may use it as a reference point for assessment of other, subsequent numbers.

It is not just numbers that give context, of course: contextual language can be an expression of the communicator's own interpretations (or what they want the audience to think about the numbers). For instance, a crowd of about 150 people could be described as 'more than 100' or 'fewer than 200'. While both descriptions are true, they can betray information about the communicator's evaluation of the size of the crowd. Similarly, phrases like 'only 20%' or 'just 5,000' convey extra information about communicators evaluations of the numbers (Teigen, 2022). This kind of information is helpful and perhaps even desirable for those wanting to guide an audience's perception – but should be treated with caution for those aiming for neutrality.

## 2.5. Using comparators

Small numbers or probabilities provide a particular challenge when trying to communicate the real sense of them. There is a temptation to reach for similar small numbers as comparators, the quintessential one being 'the chance of being struck by lightning' (less than 1 in a million chance in the US) (CDC, 2023).

The lightning comparison appears to have been a mainstay of risk communication during the pandemic, particularly in the case of describing of rare side effects of COVID-19 vaccines. A cursory review of UK newspapers on the Factiva database (factiva.com; Dow Jones) reveals the phrase "struck by lightning" appears in 136 news articles mentioning COVID-19 vaccines. We suspect the driving force behind these comparisons is an effort to make the risk of vaccine side effects *acceptable*. That is, we should tolerate the risk of vaccine side effects because we do not fret about lightning strikes in our day-to-day lives. However, these two events, vaccine side effects and lightning strikes couldn't be more different (other than their rough probability of occurring at population level).

When probabilities and outcomes (e.g. death) are equal there are still a range of extra factors which determine how we feel about, and decide on the acceptability of, different risks (see section 2.1). Comparing unrelated risks, such as vaccination side effects with lightning strikes, may not have the desired effect of facilitating accurate risk perceptions (Visschers et al., 2009). We echo the suggestion of Visschers et al. (2009) that if drawing comparisons, risks with similar characteristics, including their emotional salience, should be selected with care. Indeed, when comparing the risks and benefits of vaccine in the graphic by the Winton Centre for Risk and Evidence Communication in section 2.7, it was important to choose the correct comparator (infection leading to ICU admission versus serious adverse vaccine reaction), and when interviewing people to find what would help put their individual risk of dying from COVID-19 into context for them, the authors found that it was a comparison with *other people's risk* of dying from *COVID-19*, not *their risk* of dying from *other causes* (Freeman et al., 2021). Comparing 'within the risk' rather than with other risks therefore seems a promising strategy, but more research is needed on a case-by-case basis. While comparisons may be useful in some cases (e.g. Keller et al., 2009; Savadori et al., 2022), we urge communicators to proceed with caution.

## 2.6. Communicating uncertainty around numbers

No estimate is precise, no empirical evidence is 100% certain – uncertainty is an inherent part of science and the scientific process. Thinking about how this can be communicated and its effects on people's understanding and perception of the information at hand is therefore an important question, especially in times of crisis when uncertainty is inherent. The uncertainty around a numerical estimate or statistic has been described as statistical or 'direct' uncertainty (Kerr et al., 2022; van der Bles et al., 2019) (the counter-part to 'indirect' uncertainty, which we will introduce in section 3). For example, the statistical confidence interval around a point estimate constitutes direct uncertainty, and it can be communicated as a numerical range, visually, or through a verbal indication.

Early in the COVID-19 pandemic, little data was available so modelling estimates and statistics came with a high degree of uncertainty which was not always reflected in the presentation of the numbers. As the pandemic progressed, there was greater use of uncertainty bounds in presentations, even when the uncertainty was still high. Much research has shown that uncertainty and its communication affect how people perceive a piece of information and how they use it (Gustafson & Rice, 2020). It is therefore important to understand how different expressions of uncertainty affect an audience.

A lot of research has focused on effects on trust, with mixed findings. The majority of work seems to suggest that communicating direct uncertainty, in the form of numeric ranges around an estimate, does not undermine trust (Gustafson & Rice, 2020; van der Bles et al., 2020). Some work even suggests that communicating uncertainties can increase trust in some contexts (Howe et al., 2019; Joslyn & Demnitz, 2021; Joslyn & LeClerc, 2016). However, some work found that presenting numeric ranges can undermine trust in science and science-based policymaking (Kreps & Kriner, 2020).

The picture looks different for verbal expressions of uncertainty which are by nature less concrete and more ambiguous, for instance statements such as 'some uncertainty'. Here research has reported potential negative effects on trust and other outcome measures (Han et al., 2018; Kreps & Kriner, 2020; van der Bles et al., 2020), echoing some of the concerns relating to the use of words versus numbers when communicating statistical and empirical information generally, outlined in section 2.3.

Adequately communicating uncertainty plays a particularly important role in times of crisis such as the COVID-19 pandemic. A study on the official communication of SARS-CoV-2 test results across three countries (UK, USA, New Zealand) - which were characterized by different levels of expression of uncertainty - found that the way in which uncertainty is described (or not) did affect people's perceptions and interpretations of the results. The official UK wording early on in the pandemic did not mention the possibility of false negatives or false positives while the official New Zealand wording highlighted uncertainties more explicitly. Those participants reading the UK wording in the study were more likely to give categorical, so more definitive answers (100% or 0%) when asked for their estimate of the likelihood of infection given a test result (positive vs. negative), with the reverse being the case for those reading the New Zealand wording. These tendencies were also reflected in people's ratings on whether a symptomatic individual who tests negative should self-isolate. The proportion of participants who indicated that the symptomatic individual (who tested negative) should definitely *not* self-isolate was highest for those reading the UK wording and lowest for those reading the NZ wording (Recchia et al., 2021). Expressing no uncertainty, then, can be misleading as people might assume that information they are given is more reliable than it actually is and display unwarranted confidence in it. Apart from ethical considerations which call for disclosure and communication of uncertainty, there are also other potential long-term benefits when information changes, as we will discuss in section 3.2. As for the format in which uncertainty around numbers is best communicated, research indicates that numeric representations clearly outperform verbal expressions and should therefore be preferred.

## 2.7. Visualizing numbers

A picture is worth a thousand words. Visualisations can help us understand datasets that might otherwise be difficult to interpret (Lipkus & Hollands, 1999; Tufte, 1983). They can also be intuitive - reducing the cognitive load required when trying to understand risk and uncertainty communications and allowing patterns in data to be rapidly extracted (e.g. Padilla et al., 2022a; Pomerantz & Pristach, 1989; Scaife & Rogers, 1996). They can further transcend language barriers and be accessible across numeracy and literacy levels; essential for fast, equitable communication to all in a crisis. Some examples, which we will discuss below are given in Figure 2. However, it is important to use the right kind of visualisations in the right circumstances.

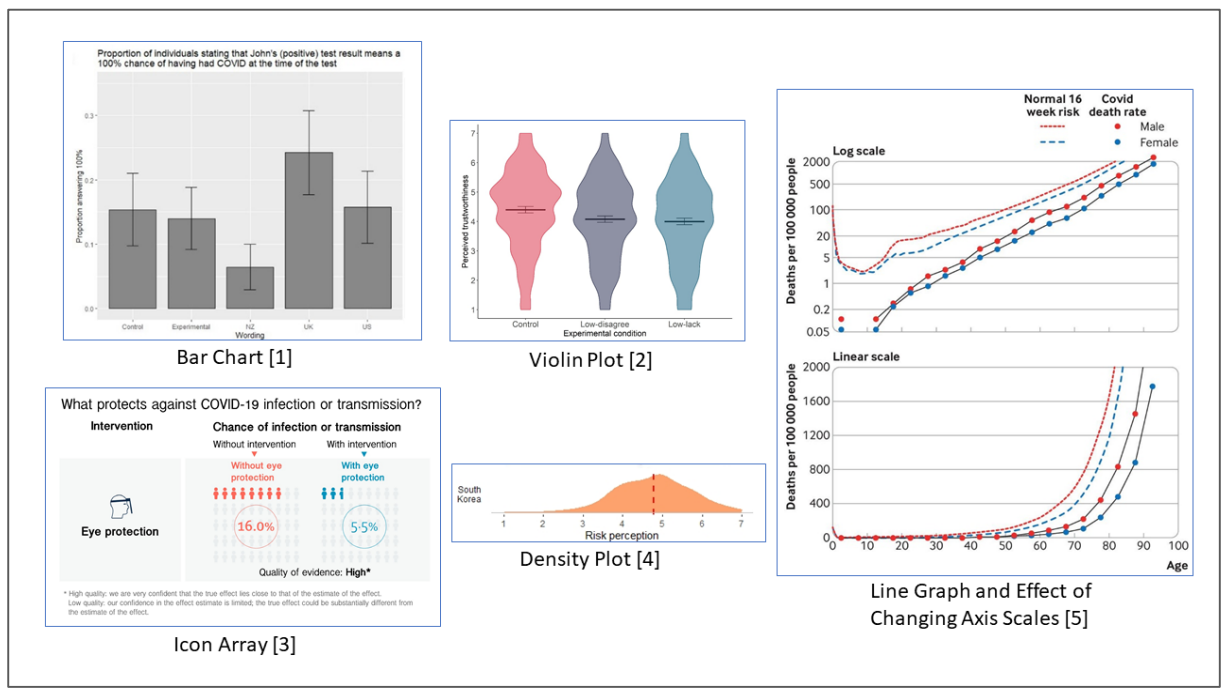

**Figure 2.** Examples of different visualisation techniques. Taken from: [1] Recchia et al. (2021), [2] Schneider et al. (2022), [3] Schneider et al. (2021), [4] Dryhurst et al. (2020), [5] Spiegelhalter (2020)

The fast-thinking that is common to mental processing of data can cause us to fall foul of biases in interpretation (Kahneman, 2011), particularly when the brain is trying to map visuals back to their original numeric value (Franconeri et al., 2021; Huff, 1954; Tufte, 1983). Poorly designed visualisations then, can lead to misperceptions of risk, misunderstanding and lack of engagement in preparedness or protective behaviours (Padilla et al., 2022a). During the COVID-19 crisis, thousands of visualisations of pandemic data were produced worldwide (Zhang et al., 2021), however few have been evaluated, and thus there is fairly limited understanding of the efficacy of these visualisations for helping audiences understand pandemic risks, or their effects on risk perceptions and resultant behaviours (Padilla et al., 2022a).

A notable exception is Padilla et al. (2022a), who evaluated the impact of all 34 types of COVID-19 data visualisation techniques available on the CDC's website at the time of publication on people's risk perception of the same COVID-19 mortality data. They demonstrated that participants' risk perceptions of their own and others' risks from COVID-19 consistently increased after seeing visualisations of mortality data represented on cumulative scales, whereas the effects of weekly incidence scales on risk perceptions were much more variable. They also showed that the way in which uncertainty in COVID-19 forecasts was visualised affected risk perceptions, with visualisations displaying six or more models increasing estimates of COVID-19 risk more than other formats tested, including those showing a gradient depiction of forecast uncertainty and those showing 60% or 95% confidence interval summary statistics.

Even commonly-used visualisations that might traditionally be seen as easily understood (e.g. bar charts, uncertainty cones, error bars) have been shown to be misinterpreted when evaluated empirically (Correll & Gleicher, 2014; Newman & Scholl, 2012; Padilla et al., 2017; Ruginski et al., 2016). For instance, the variation depicted by error bars (e.g. 95% confidence intervals, standard errors) can be misinterpreted, resulting in misperceptions about the size of statistical effects (e.g. Hofman et al., 2020); an effect to which not even experts are immune (Belia et al., 2005).

When designing a visualisation, it is important to understand the decisions people might be taking based on it, and hence use the visualisation technique that best supports these decisions (ideally having empirically evaluated that this is indeed the case for the specific visualisation used). Visualisations of expected frequencies, such as icon arrays, for example, can be very useful in helping people intuitively understand a probability, and comparing two probabilities. They use coloured dots, lines or icons to visually represent the expected number of events out of a total number (e.g. 100 dots of which 5 are red, representing an expected frequency of 5 out of 100). There is much evidence that icon arrays are effective for communicating healthcare treatment outcomes (e.g. Galesic et al., 2009; Garcia-Retamero et al., 2010), and can be particularly useful for those with lower numeracy skills in understanding probabilistic information (e.g. Galesic et al., 2009; Garcia-Retamero & Galesic, 2009; Hawley et al., 2008). Figure 3 shows a graphic from the Winton Centre for Risk and Evidence Communication which uses an icon array approach to visualizing harms and benefits of the Astra-Zeneca COVID-19 vaccine. An earlier version of this graphic was utilized by the Deputy Chief Medical Officer Jonathan Van Tam in his briefing on 7 April 2021 explaining the decision-making process around the UK vaccination programme.

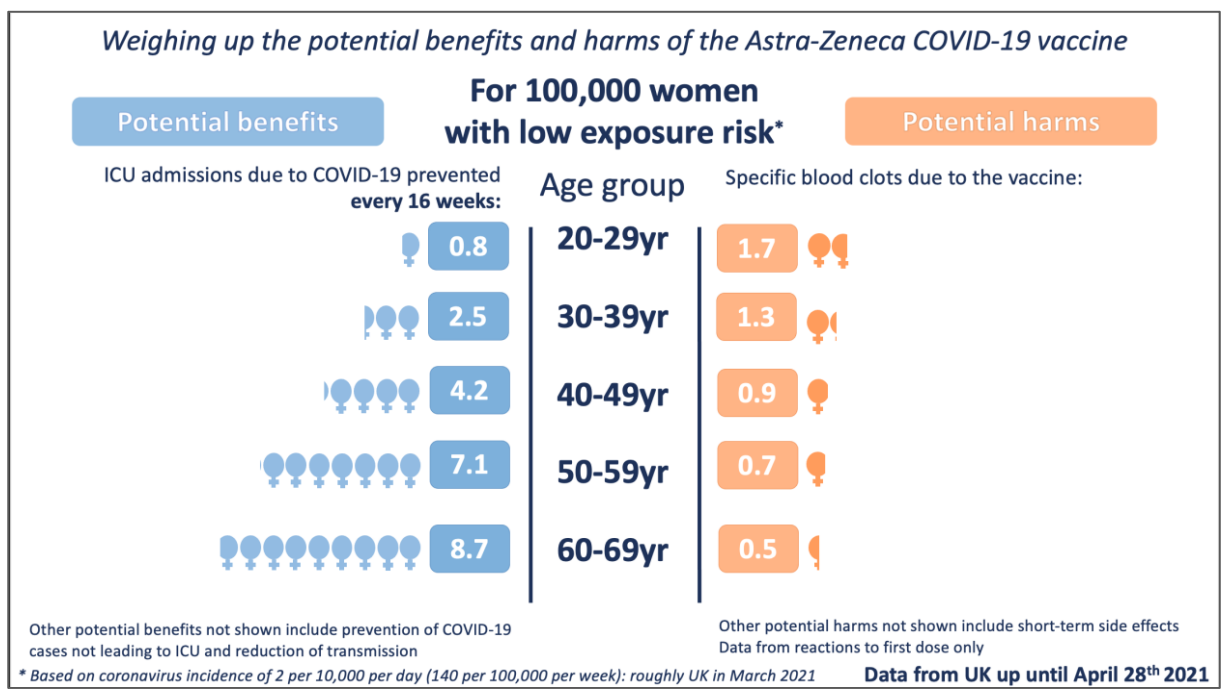

**Figure 3.** Winton Centre for Risk and Evidence Communication graphic on harms and benefits of Astra-Zeneca COVID-19 vaccine. An icon array would typically include an explicit visualization of the denominator. The icon array shown here is an alternative variation, in which the denominator is not shown, as it is too large. This graphic is one (low exposure for women) of a set of six made for policy decision-makers deciding the UK's national vaccine policy after the first signs of potential harms from the AstraZeneca COVID-19 vaccine emerged. It uses a 'butterfly' style graphic design, with the potential benefits and potential harms of the vaccination being weighed up on each side of the butterfly. For these policy decision-makers, it was important to stratify the data so that they could take into account the potential impacts of any policy decision on different subgroups of the population. For example, the effects on men versus women, those of different ages, and those exposed to different incidence levels of the virus were all illustrated across the 6 slides. Using these visualisations alongside their knowledge of other important factors (such as the availability of alternative vaccines) the policy-makers were able to form a policy based on the current evidence, and then used these graphics to explain their decision to the public. For individual members of the public, these graphics also allowed them to look at the potential benefits and harms to them – dependent on the major risk factors of age, sex and exposure.

When illustrating change over time, a line plot is a common format. It should be noted at this point that line plots can vary considerably, especially in their axes, for example linear versus logarithmic scales. In a review of over 600 visualisations of COVID-19 data (Zhang et al., 2021) found that a line chart showing how different COVID-19 summary metrics changed over time was the most common format used by news media, academics and government agencies (although these varied in whether they showed cumulative or non-cumulative data, or whether they presented data linearly or on a non-linear scale (Padilla et al., 2022a)). Indeed, the scale of a line chart can change its impacts on the audience. Romano et al. (2020) evaluated line charts used in COVID-19 data communication and found that people viewing a line chart with a linear scale showed higher comprehension than those viewing a non-linear version (in line with other studies on comprehension of linear versus non-linear scaling). They also found these participants were more worried about COVID-19 after viewing the linear chart.

If the uncertainty around estimates is important to communicate for a decision-maker then different visualisations are needed. Visualising the underlying probability distribution and mapping probability visually to the width of this distribution (e.g. via density plots; violin plots) can help people intuitively understand not just the most likely value but also the shape of the distribution and extent of variability (Franconeri et al., 2021). Such distributional visualisations have been shown to improve accuracy and reduce mental effort over text-based expressions and equivalent summary statistics such as means and confidence intervals (Castro et al., 2022; Correll & Gleicher, 2014).

For visualising geospatial or temporal uncertainty information, an ensemble plot might be more helpful. These have been found to be better at communicating uncertainty in hurricane forecasts, for example, than the usual "cone of uncertainty", originally developed by the National Hurricane Centre. The 'cone' represents spatial confidence intervals around the most likely forecast (a 66% confidence interval in the case of hurricane forecasts), but people thought the cone represented the shape of the forecasted storm, growing over time (Padilla et al., 2017; Ruginski et al., 2016). An ensemble plot, by contrast, displays a selection of representative ensemble members derived from runs of individual forecasting models on a common Cartesian coordinate plane (Brodlie et al., 2012; Harris, 2000; Padilla et al., 2017; Potter et al., 2009). Compared to the cone of uncertainty, ensemble plots improve comprehension of hurricane forecast information and dispel misunderstandings that the size of the display represents the size of the storm (Liu et al., 2019). Zhang et al. (2021) found that 29% of the 600 COVID-19 visualisations they identified visualised COVID-19 related information as ensembles of possible scenarios or model runs. As with any approach to communicating risk and uncertainty however there are some downsides of ensemble plots. For example, they can cause viewers to overestimate the likelihood of a particular hurricane if it passes over a location that carries significant meaning, such as their hometown (Padilla et al., 2017; Padilla et al., 2020). There are other visualization formats not covered here which can be useful depending on the context (Franconeri et al., 2021; Hullman et al., 2015; Kale et al., 2021; Padilla et al., 2022b). All formats come with their specific advantages and drawbacks, and their use should be carefully considered, taking into account the specific context, audience, and purpose.

There are also many examples of visualisations being deliberately used to persuade or even mislead. Common techniques include truncating axes so the data run over a smaller range, making differences appear larger (Correll et al., 2020; Pandey et al., 2015), and inverting axes so they run from larger to smaller values (Franconeri et al., 2021; Pandey et al., 2015).

Colour is an important component of good visualisation design, helping users rapidly grasp the gist of communications and intuit properties such as data groupings and quantities such as magnitude. Indeed, colour intensity is one of the limited visual channels that can effectively communicate magnitude to an audience (Franconeri et al., 2021). Choosing colours for groupings that are more distinct in perceptual space (for example red and blue by contrast to red and orange) makes distinctions between data groupings easier to draw (Franconeri et al., 2021). Colour is also a useful cue to pair with spatial proximity; in a scatter plot, if the colour coded plotted points are not proximally grouped in space, the viewer can intuit that the distinction between those two groups as they relate to the x and y axis is limited.

The use and interpretation of colour however, is not neutral; different presentations can lead to biases in interpretation. Indeed, some colours carry intuitive meanings, such as red for danger or blue for water (Bostrom et al., 2008; Hagemeier-Klose & Wagner, 2009; Thompson et al., 2015). Whilst in some instances these automatic associations can aid comprehension by enabling rapid intuition of meaning, in others, they may also cause communications to be seen to convey information they do not actually contain. This might be of considerable import in times of crises where such miscommunication can lead not only to loss of trust, but also to loss of life. Choice of colour can also be used with deliberate intent to drive a particular message, or to mislead. It is also worth noting that such meanings can be culturally and context-specific and so the choice of colour must always be made carefully.

Interestingly, some of the ambiguities of colour can be turned to good use when communicating uncertainty information. Value-suppressing uncertainty palettes are a fairly new technique design to convey uncertainty in the information being communicated by giving the viewer an implicit feeling of uncertainty (Correll et al., 2018). Such palettes link values to different colour hues and uncertainty to colour saturation and lightness. When colours are less saturated and lighter they become harder to distinguish, making it more difficult for the viewer to compare between such values. This conveys an implicit sense of uncertainty to the viewer, with the most uncertain values in such a palette appearing the same shade of grey (Correll et al., 2018; Franconeri et al., 2021).

It is important to note that colour vision impairments are estimated to affect approximately 4% of the population (Olson & Brewer, 1997), which has serious implications for the accessibility of visualisations. We thus recommend communicators use colour palettes that have been designed with colourblind users in mind (e.g. colorbrewer.org).

For an extensive review of data visualisation design including the use of colour and other design features we highly recommend Franconeri et al. (2021).

## 3. Communication beyond numbers: Evidence quality, change, and transparency

The communication of empirical evidence goes beyond the presentation of numbers, such as the outputs of modelling processes, to encompass other elements associated with the evidence. Such aspects include the confidence in and quality of the evidence base, the question of dealing with and reacting to a changing evidence base over time, repeated decision making based on an evidence base, and the degree of transparency and openness about uncertainties. In this section we shed light on some of these broader considerations which might not immediately come to mind when thinking of the communication of statistics, but which are nevertheless crucially important. We will provide some pointers to relevant current literature in order to highlight avenues for further reading.

## 3.1. Quality of evidence communication

The communication of scientific estimates and statistics does not only involve communicating numbers such as the outputs of mathematical modelling, but - importantly – also information about the underlying evidence base. Information on the evidence base allows us to put a statistic into perspective and give it context. For instance, a medical treatment that appears to shows high effectiveness can mean very different things if the underlying evidence on which the effectiveness estimate is based is on several large randomised, controlled trials, compared to when it is based only on a few case studies. It has been argued that the communication of certainty of evidence matters not just generally but especially in situations of "emergencies and urgencies" (Schünemann et al., 2020, p. 202).

For example, a vaccine report from the UK Health Security Agency during the pandemic presented infection rates in vaccinated versus unvaccinated people. However, the estimated population of unvaccinated individuals in the UK was drawn from a data source which overestimated the total population of unvaccinated individuals. In turn, this led to underestimation of the SARS-CoV-2 infection rate among unvaccinated, ultimately leading to the impression that vaccination may *increase* one's risk of SARS-CoV-2 infection. Despite the limitations of the data being noted several pages later in the report, the results were seized upon as evidence of the vaccines' ineffectiveness (Gregory, 2021; Jones, 2021; Reuters Fact Check, 2021; UK Health Security Agency, 2021). Could this have been avoided if the quality of the evidence base - including the methodological problems in calculating the rates - had been communicated differently?

Sometimes exact numbers are not available because of a lack of reliable evidence. It can be useful however, and sometimes necessary, to communicate the gist information. In such cases using colour to represent the information can be a helpful tool. See the sidebar for a visualization of expert estimates of SARS-CoV-2 transmission which, despite avoiding numbers (because they were too imprecise given the low quality of evidence), still communicated current expert understanding to the extent that they could be used in decision-making.

The communication of an assessment of the evidence base is commonly referred to as communicating 'quality' or 'certainty of evidence', 'confidence', or 'indirect uncertainty' (with direct uncertainty being the uncertainty surrounding numerical estimates, such as confidence intervals, see section 2.6.) (Schneider et al., 2021; Schneider et al., 2022; van der Bles et al., 2019). Quantifying indirect uncertainty is not an easy task, as quality assessments are qualitative, and more subjective in nature (Schneider et al., 2021; van der Bles et al., 2019). Spiegelhalter (2017) refers to 'unmodeled uncertainty' as the uncertainty that stems from the quality of the evidence base which is not modelled as part of statistical analysis.

Spiegelhalter (2017) and Spiegelhalter and Riesch (2011) reviewed approaches for communicating limitations in scientific understanding and confidence in the analytic process. Summary formats to communicate strength or quality of evidence, such as verbal or numeric star-rating scales, are increasingly used in practice. Prominent examples are the GRADE (Grading of Recommendations Assessment, Development and Evaluation) system from medicine (Balshem et al., 2011; Guyatt et al., 2008a; Guyatt et al., 2008b) which assigns one of four quality levels (later renamed 'certainty levels' (Hultcrantz et al., 2017)): *High, Moderate, Low, Very low;* or the IPCC's (Intergovernmental Panel on Climate Change) summary qualifiers for evidence (*limited, medium, robust*) and agreement (*low, medium, high*), which are synthesized into confidence ratings (*very low, low, medium, high, very high*) (IPCC, 2022). These quality labels are, of course, the result of a quality assessment of the evidence base, which may follow a set process (see for instance ROBINS-I tool, Schünemann et al., 2019; Sterne et al., 2016). Such assessments include questions around the types of studies that form part of the evidence base (e.g. randomized controlled trials versus observational data), or whether there are any methodological shortcomings in the evidence base (e.g. whether an experiment was adequately blinded).

A lack of evidence or evidence from many conflicting studies could both lead to a 'low' quality qualifier, despite the qualitative difference between the two. In times of crisis, such as the COVID-19 pandemic, data tends to be scarce, therefore alternative ways of communicating the quality of the evidence might be needed. One route, for which we encourage further empirical testing, could be to highlight the reason for a low quality label, rather than just stating low quality.

Research consistently shows that the communication of evidence quality matters to audiences, and should be taken seriously (Brick et al., 2020; Brick & Freeman, 2021; Schneider et al., 2021; Schneider et al., 2022). However, exactly *how* evidence quality is best communicated is likely to be context dependent, and – like all communication – should be co-designed and evaluated with the intended audience.

Whilst a sizeable body of research exists on the effects of communicating direct uncertainty (section 2.6) the effects of communicating quality of evidence is largely unexplored. Some recent work has investigated summary formats such as GRADE's short descriptive labels ('low quality', 'high quality') and its effects on measures of trust in the information itself and in the producers of the information, as well as on how people use the information subsequently. In a set of experimental studies communicating the effectiveness around the use of eye protection for prevention of SARS-Cov-2 transmission (Schneider et al., 2021) and communicating the SARS-CoV-2 case fatality rate (Schneider et al., 2022), the authors found that people reacted to quality information in predictable ways: when the evidence was described as high quality participants trusted the information and producers more, perceived eye protection to be more effective, indicated higher intentions to wear eye protection themselves, and were more likely to use the case fatality rate in their decision making, compared to when the evidence was described as of low quality. However, when *no* indicator of evidence quality was provided, levels of trust, perceived effectiveness, and use in decision making were not significantly different from levels when the evidence was described as of *high* quality. This suggests that when evidence quality is not disclosed, people tend to assume high quality. Regardless of the psychological process leading to it, this presents a potential ethical dilemma when evidence quality is low: disclosing it might lower people's response to the findings, but not disclosing it may mislead them into thinking it is likely to be high quality. Additionally, when evidence quality was ambiguous (i.e. it was stated that it was uncertain and could be high or low), people reacted to it as if it was low (Schneider et al., 2022). These findings emphasise the importance of evidence quality in people's perceptions of an estimate or claim and related decision making. Other research in the climate domain has reported comparable findings (Howe et al., 2019).

This also spurs the question of how decisions should be made when the evidence base is of low quality. This is especially relevant in times of crisis when it is not always possible to achieve higher levels of evidence quality. During the COVID-19 pandemic information was scarce and rapidly changing in the early stages, and uncertainties were high, yet decisions had to be made on the available evidence base. An approach, which was for instance taken in the UK by the Scientific Advisory Group in Emergencies (SAGE), was to proceed with decision making while openly disclosing low evidence quality. In a SAGE meeting in early 2020 the group agreed that confidence in current modelling conclusions was low and that further review was needed, however, that given the inability to generate more evidence and more robust conclusions in the short run, policy decisions should nevertheless be based on the currently available modelling outcomes (Evans, 2021, p. 63).

The key is that in cases where estimates which are fairly uncertain are communicated, the uncertainty should be adequately expressed. SAGE, for example, qualifies its information not just with an expression of confidence but also highlights the uncertainty around the effect size; for instance: "It is almost certain that Delta variant (B.1.617.2) has a significant growth rate advantage over Alpha variant (B.1.1.7) (high confidence), though there remains considerable uncertainty around the extent of this advantage." (UK Government, 2021, point 3).

Communicating evidence quality is important, but complex. It effects audiences, in particular when highlighting low quality, and communicators should be aware of this. How exactly low quality of evidence might be communicated should be assessed in context. For instance, if assigning and communicating a label of 'low' quality might lead people to take away that the science is bad (rather than for instance there being a lack of evidence), then this misunderstanding should be avoided by communicating more clearly. Adequately communicating evidence quality is not an easy task and more research is needed to help communicators better understand the effects of various approaches.

**Sidebar:** Communicating evidence when there are no precise numbers

Visualisation can be particularly useful in the case where knowledge is so uncertain that it would be inappropriate to assign a point estimate or even a range – and yet enough is known to be able to aid decision-making if only it could be communicated in a useful way.

An example of this is in communicating transmission pathways for SARS-CoV-2 (Rutter et al., 2021), and how much different mitigation measures (such as masks, social distancing or ventilation) might reduce them. Putting precise numbers on the percentage of infectious virus that might reach one person from another, infected, person in different circumstances is impossible. However, experts have a 'sense' for the magnitude of that percentage and conveying that to policy-makers and individual members of the public is important.

One way of doing this is to use colour intensity to represent quantity of virus in a diagram illustrating the potential pathways (with and without different mitigations). Since people instinctively recognise more intense colours as representing higher magnitudes (Gaspar-Escribano & Iturrioz, 2011), but the colours cannot be taken to be representing a precise number, they can give an accurate representation of the current state of knowledge. This allows people to make decisions based on that knowledge (Stahl-Timmins, 2021).

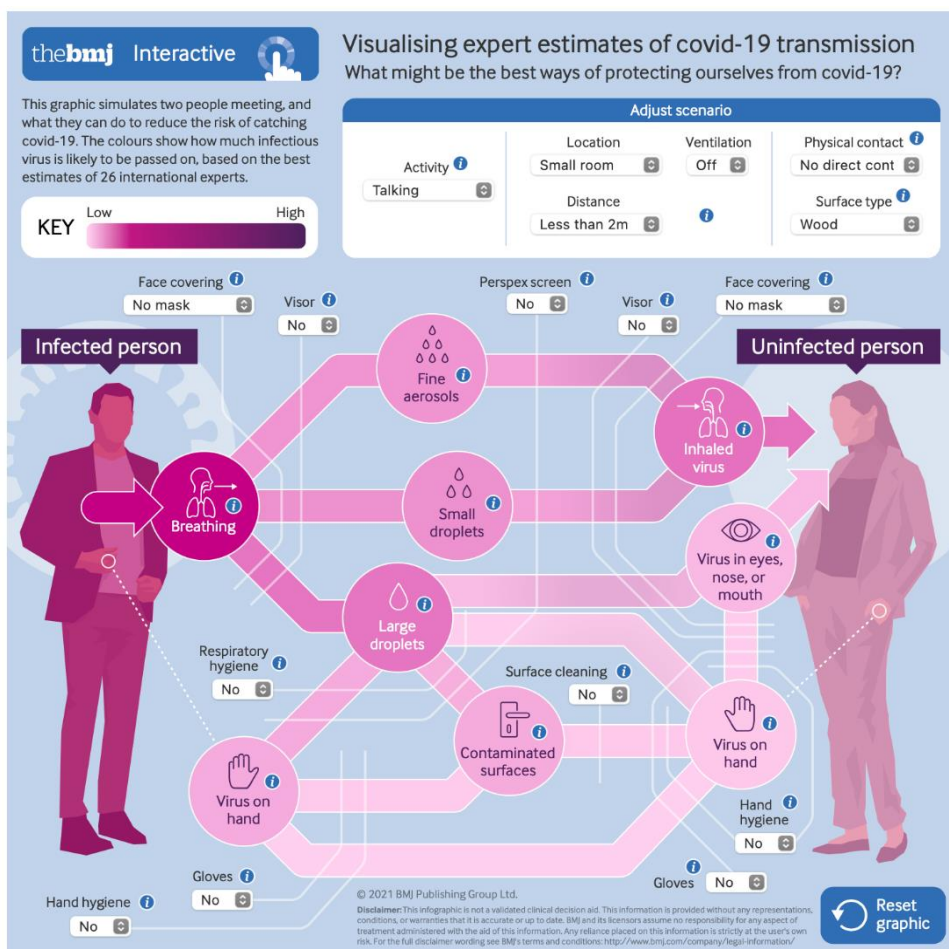

Screenshot of interactive figure taken from: https://doi.org/10.1136/bmj-2021-065312

## 3.2. What if information changes?

As could be seen during the COVID-19 pandemic, available information, including numbers on deaths and infections, along with updated information on infection pathways, efficient health protective measures and others, changed on a daily basis. The pandemic has seen advice change over its course, sometimes drastically, such as the WHO's early recommendations of not using facial masks and its later opposite recommendations of using them, feeding into a long-lasting public debate over the effectiveness of masks (Peeples, 2021; Taylor & Asmundson, 2021; https://www.wired.com/story/how-masks-went-from-dont-wear-to-must-have/). This sparks the question of how best to signal to audiences that change should be expected. Recent work showed that communicating uncertainties about COVID-19 vaccine effectiveness could buffer against negative effects of changing information: those who were presented with certain (compared to uncertain) effectiveness information showed a greater loss in vaccination intentions and trust in government after having received conflicting effectiveness information (Batteux et al., 2022). Other work reports similar findings: while downplaying uncertainty may be beneficial in the short term (communicating uncertainty can erode trust in some contexts), it can have adverse effects in the long run when information changes (Kreps & Kriner, 2020). Insights from this line of research suggests that being upfront about uncertainties and not communicating with false certainty might be beneficial in the longer term. These findings support the communication strategy adopted and advocated for by Lord Krebs, head of the UK Food Standards Agency through several crises in the 2000s. Key elements are to say what you know, but also what you don't know; what you are doing to find out more; what the advice is in the meantime, and that advice might change as new evidence arises (Blastland et al., 2020; Champkin, 2013). Similar points have been made by other groups (e.g. Hyland-Wood et al., 2021; Petersen et al., 2021; Veit et al., 2021; Vraga & Jacobsen, 2020).

## 3.3. Does transparency about uncertainty undermine trust?

One of the concerns often expressed by experts about being transparent about statistical uncertainties and the overall quality of evidence is that doing so may undermine their credibility (Fischhoff, 2012). However, empirical evidence suggests that clear communication of numeric uncertainty, such as expressed by using a range rather than a point estimate, does not undermine trust in communicators or scientists more generally (Gustafson & Rice, 2020; van der Bles et al., 2020). An exception to this may be when uncertainty intervals are very large. Kreps and Kriner (2020) report that people are less trusting of scientists after reading COVID-19 death projections presented as large range, compared to a point estimate with no uncertainty.

Blastland et al. (2020), provide a broad framework for balanced and transparent communication, distilled as 'five rules for evidence communication' (Table 3).

**Table 3.** Overview of "five rules for evidence communication" from Blastland et al. (2020), adapted from Kerr et al. (2022)

| Recommendation | Description |
| --- | --- |
| Inform, not persuade | An overarching recommendation to communicate with the aim of informing the decision-maker's choice, rather than pushing them towards a given option. |
| Offer balance, not false balance | Be clear about the benefits and costs or risks associated with decision options while acknowledging weight of evidence. |
| Disclose uncertainties | Clearly describe uncertainties around the evidence presented. |
| State evidence quality | Provide information about the quality of the evidence drawn upon. |
| Inoculate against misinformation | Identify and pre-empt circulating misinformation or misperceptions about the topic. |

These recommendations were based on a combination of empirical findings about individual elements and professional experience but had not--at the time--been empirically tested overall. Kerr et al. (2022) experimentally investigated the effect of applying these rules on trust in communications and their sources. Using communications relating to COVID-19 vaccines and nuclear power they found that versions edited to incorporate Blastland et al.'s (2020) 'five rules' were, on average, considered as trustworthy as (in the case of COVID-19 vaccines), or more trustworthy (in the case of nuclear power) than the originals.

## 3.4. Making repeated decisions

In many crises, decisions are made over short time frames and are not retaken repeatedly. In most crises the UK government SAGE mechanism meets once or at most a few times in response. During the COVID-19 pandemic, however, SAGE met repeatedly to review evidence and decisions since new scientific insights emerged continuously and over a long period of time. Making repeated decisions in a changing informational environment, such as consecutive decisions over whether to go into lockdown or not, can be challenging. Different audiences might expect different approaches to how these decisions are taken and different types of biases might colour these preferences. Understanding some of the biases at play might help to understand audience reactions.

An approach which combines learning from past experience with reacting to new incoming information is Bayesian updating (McCann, 2020). It combines existing estimates – for instance on the probability of uncertain outcomes - with an assessment of the strength of new evidence. This helps to improve the accuracy of probability estimates, supporting better informed decisions. It can also help to foster open-mindedness and lower effects of cognitive biases. Experts during the pandemic, such as Sir Patrick Vallance in the UK, have consistently tried to push for an updating approach, highlighting that evidence would be reviewed in light of new information (e.g. UK Science and Technology Select Committee Oral Evidence, 2021).

This can then of course lead to 'U-turns' if the evidence substantially changes, which might not be viewed favourably by all audiences. Research suggests that experience, such as having lived through a decision and having experienced its consequences, is a powerful component to decision making. For instance, research using a binary choice task in an organizational behaviour environment found that decision makers rely heavily on experience even when there is descriptive information available (Lejarraga & Gonzalez, 2011). People's lived experiences can influence how information and decisions are viewed.

Additionally, people may exhibit inertia or status-quo bias: a tendency to stick with a previously taken course of action and a reluctance to react to disconfirming information, a psychological phenomenon documented across domains (e.g. management, medicine) (Alós-Ferrer et al., 2016; Gal, 2006; Ritov & Baron, 1992; Samuelson & Zeckhauser, 1988; Suri et al., 2013). This may help to explain why 'U-turns' might evoke negative reactions with some audiences. We recommend clear, transparent communication around the reasons for a course of action, disclosure of uncertainties and acknowledgement that the evidence might change in the future, as discussed in sections 3.2. and 3.3.

# 4. Conclusion and recommendations

## 4.1. Summary

In this review we examined considerations for communicating empirical evidence in times of crisis. Some important take-away points are:
1. *Communication aims matter.*
   As a communicator it is important to be aware of the communication goal. Is it to inform or to persuade? If the goal is to inform, then uncertainties should be disclosed and well communicated, along with the quality of the underlying evidence. The way information is communicated might look very different if the aim is to persuade versus inform.
2. *Communicating evidence quality affects how people perceive and use the information.*
   There are ethical considerations around the disclosure of evidence quality, especially when it is low. In times of crisis, decisions must be made on incomplete evidence or evidence of poor quality. Communicating this clearly and as intelligibly as possible is key.
3. *Transparent evidence communication may be critical for retaining trust in the long run.*
   While there may be short-term benefits to omitting uncertainties, communicating with false or inadequate certainty can have detrimental downstream effects, for instance when the situation changes and recommendations need to be revised.
4. *Emotions play a big role for how people deal with and react to information.*

Humans are not purely rational machines. How we feel about situations and information affects decision making and behaviour. As a communicator it is important to be aware of the affective impact of communications and downstream effects on how people process the presented information.

5. *Format and the way in which statistical information is presented affects understanding and perception.*
Whether empirical information is communicated as numbers or words, whether comparators are used or not, whether percentages are chosen over frequencies, the way in which uncertainty is visualized and other factors all play a role in how people understand the information presented and react to it. An awareness of these factors is key.

## 4.2. Points to keep in mind when communicating statistics and empirical evidence

We hope this review provides useful insights for communicators of statistical information and empirical evidence, especially when asked to communicate in times of crisis and heightened uncertainty and change. Below we offer several pointers, based on the research we have presented in this review. We see these as adding to and building on those made by Blastland et al. (2020), Spiegelhalter (2017) and McConway and Spiegelhalter (2021). We caution the reader that this is not an exhaustive list of definitive recommendations for how to communicate but rather a starting point for consideration. We hope that the references in the review provide considerable further reading and opportunities for engaging with the relevant literature and ongoing research. We'd also like to note that not all points in the list might be relevant to every reader. For instance, the challenge around sticking to one's expertise when pressured for information might be particularly pertinent for experts who find themselves in a media-facing communication context but not others who regularly communicate a wide range of information in such contexts.

I. **Remember that numbers aren't neutral**
   - **Take care with the presentation** of empirical evidence and its uncertainties, including the use of visuals.
   - **Be relevant** by using carefully chosen comparisons.
   - **Be aware** of lived experiences and how they naturally affect people's perceptions and decisions.
II. **Remember your audience is trying to make decisions**
   - **Know your audience** – provide the information that you would want to know if you were them.
   - **Don't get lost in the numbers** – give the all the appropriate information but no more.
III. **Remember to be trustworthy**
   - **Be open and transparent** when informing people: uncertainty is not a weakness.
   - **Know one's own expertise** (and stick to it).
   - **It's OK to be good enough** – there's an art as well as a science to communication and there is no perfect way to do it.

# Disclosure Statement

# Acknowledgements

We would like to thank David Spiegelhalter, Alexandra Freeman and Duncan Cook for insightful comments on an early draft of this review. We would also like to acknowledge the funding received through the David and Claudia Harding Foundation.

Champkin, J. (2013). Lord Krebs. *Significance*, *10*(5), 23–29. https://doi.org/https://doi.org/10.1111/j.1740-9713.2013.00694.x

Cologna, V., & Siegrist, M. (2020). The role of trust for climate change mitigation and adaptation behaviour: A meta-analysis. *Journal of Environmental Psychology*, *69*, 101428. https://doi.org/https://doi.org/10.1016/j.jenvp.2020.101428

Correll, M., Bertini, E., & Franconeri, S. (2020). Truncating the Y-Axis: Threat or Menace? *Proceedings of the 2020 CHI Conference on Human Factors in Computing Systems*, 1–12. https://doi.org/10.1145/3313831.3376222

Correll, M., & Gleicher, M. (2014). Error Bars Considered Harmful: Exploring Alternate Encodings for Mean and Error. *IEEE Transactions on Visualization and Computer Graphics*, *20*(12), 2142–2151. https://doi.org/10.1109/TVCG.2014.2346298

Correll, M., Moritz, D., & Heer, J. (2018). Value-Suppressing Uncertainty Palettes. *Proceedings of the 2018 CHI Conference on Human Factors in Computing Systems*, 1–11.

Cuite, C. L., Weinstein, N. D., Emmons, K., & Colditz, G. (2008). A test of numeric formats for communicating risk probabilities. *Medical Decision Making*, *28*(3), 377–384. https://doi.org/10.1177/0272989X08315246

Denes-Raj, V., Epstein, S., & Cole, J. (1995). The Generality of the Ratio-Bias Phenomenon. *Personality and Social Psychology Bulletin*, *21*(10), 1083–1092.

Dryhurst, S., Schneider, C. R., Kerr, J., Freeman, A. L. J., Recchia, G., van der Bles, A. M., Spiegelhalter, D., & van der Linden, S. (2020). Risk perceptions of COVID-19 around the world. *Journal of Risk Research*, 1466–4461. https://doi.org/10.1080/13669877.2020.1758193

Evans, R. (2021). SAGE advice and political decision-making: 'Following the science' in times of epistemic uncertainty. *Social Studies of Science*, *52*(1), 53–78. https://doi.org/10.1177/03063127211062586

Falbén, J. K., Golubickis, M., Tamulaitis, S., Caughey, S., Tsamadi, D., Persson, L. M., Svensson, S. L., Sahraie, A., & Macrae, C. N. (2020). Self-relevance enhances evidence gathering during decision-making. *Acta Psychologica*, *209*, 103122. https://doi.org/https://doi.org/10.1016/j.actpsy.2020.103122

Finucane, M. L., Alhakami, A. L. I., Slovic, P., & Johnson, S. M. (2000). The affect heuristic in judgments of risks and benefits. *Journal of Behavioral Decision Making*, *13*(1), 1–17. https://doi.org/10.1002/(SICI)1099-0771(200001/03)13:1<1::AID-BDM333>3.0.CO;2-S

Fischhoff, B. (2012). Communicating Uncertainty: Fulfilling the Duty to Inform. *Issues in Science and Technology*, *28*(4), 63–70.

Fountain, H. (2011). Trial over earthquake in Italy puts focus on probability and panic. *New York Times*, D3-L.

Franconeri, S. L., Padilla, L. M., Shah, P., Zacks, J. M., & Hullman, J. (2021). The Science of Visual Data Communication: What Works. *Psychological Science in the Public Interest*, *22*(3), 110–161. https://doi.org/10.1177/15291006211051956

Freeman, A. L. J., Kerr, J., Recchia, G., Schneider, C. R., Lawrence, A. C. E., Finikarides, L., Luoni, G., Dryhurst, S., & Spiegelhalter, D. (2021). Communicating personalized risks from COVID-19: Guidelines from an empirical study. In *Royal Society Open Science* (Vol. 8, Issue 4). https://doi.org/10.1098/rsos.201721

Gal, D. (2006). A psychological law of inertia and the illusion of loss aversion. *Judgment and Decision Making*, *1*(1), 23–32. https://doi.org/DOI: 10.1017/S1930297500000322

Galesic, M., Garcia-Retamero, R., & Gigerenzer, G. (2009). Using Icon Arrays to Communicate Medical Risks: Overcoming Low Numeracy. *Health Psychology*, *28*(2), 210–216. https://doi.org/10.1037/a0014474

Garcia-Retamero, R., & Galesic, M. (2009). Communicating treatment risk reduction to people with low numeracy skills: A cross-cultural comparison. *American Journal of Public Health*, *99*(12), 2196–2202. https://doi.org/10.2105/AJPH.2009.160234

Garcia-Retamero, R., Galesic, M., & Gigerenzer, G. (2010). Do icon arrays help reduce denominator neglect? *Medical Decision Making*, *30*(6), 672–684. https://doi.org/10.1177/0272989X10369000
27